\documentclass[superscriptaddress, twocolumn, aps, pra, 10pt]{revtex4-1}
\usepackage{txfonts}
\usepackage{graphicx}
\begin{document}
\title{Absolute frequency measurement with uncertainty below $1\times 10^{-15}$ using International Atomic Time }
\author{Hidekazu Hachisu}
\affiliation{National Institute of Information and Communications Technology, Koganei, Tokyo 184-8795, Japan}
\author{G\'{e}rard Petit}
\affiliation{International Bureau of Weights and Measures (BIPM), Pavillon de Breteuil, 92312 S\'evres, France}
\author{Tetsuya Ido}
\email{email: ido@nict.go.jp}
\affiliation{National Institute of Information and Communications Technology, Koganei, Tokyo 184-8795, Japan}

\date{\today}
\begin{abstract}
The absolute frequency of the $^{87}{\rm Sr}$ clock transition measured in 2015 \cite{jjap} was reevaluated using an improved frequency link to the SI second. The scale interval of International Atomic Time (TAI) that we used as the reference was calibrated for an evaluation interval of five days instead of the conventional interval of one month which is regularly employed in Circular T. The calibration on a five-day basis removed the uncertainty in assimilating the TAI scale of the five-day mean to that of the one-month mean. The reevaluation resulted in the total uncertainty of $10^{-16}$ level for the first time without local cesium fountains. Since there are presumably no correlations among systematic shifts of cesium fountains worldwide, the measurement is not limited by the systematic uncertainty of a specific primary frequency standard.
\end{abstract}

\maketitle
\section{Introduction}
\label{intro}
The progress of optical frequency standards has recently been further accelerated to realize the accuracy at the $10^{-18}$ level \cite{NIST,RIKEN,JILA,PTB}.  This accuracy is nearly two orders of magnitude superior to the state-of-the-art microwave standards that realize the SI second. %It is not satisfactory that the SI second is no longer based on a transition realized with the smallest uncertainty.
Thus, the community of time and frequency metrology has initiated discussion toward the redefinition of the second \cite{Riehle,Gill_FSM}. As the discussion has proceeded, however, it has been recognized that the requirement for the redefinition will not be  fulfilled soon, possibly delaying the redefinition by nearly a decade; until then, we need to maintain traceability to the current SI second. While the most straightforward method of accessing the SI second is to develop a cesium (Cs) fountain, the difficulty of building state-of-the-art standards limits the availability.

The frequency link to International Atomic Time (TAI) is known to be an alternative means of accessing the SI second \cite{nrc2012,kriss2013,aist2014,wipm2015,nict2011}.  In particular, temporally  distributed operations of an optical clock over a five-day TAI grid have reduced the possibility that the measurement is misled by the temporal frequency fluctuation of a local flywheel oscillator \cite{jjap}. While the up time ratio was only 11\% over the five-day campaign, the dead time uncertainty between the optical clock and the local flywheel oscillator was reduced to $2.7 \times 10^{-16}$. Thanks to these efforts, the total uncertainty of the absolute frequency has resulted in $1.1\times 10^{-15}$ \cite{jjap}, which is already less than the systematic uncertainty of some Cs fountain standards that are recognized by the International Bureau of Weights and Measures (BIPM) as the primary frequency standards (PFS).

TAI is computed by BIPM from the clock data provided by more than 400 commercial atomic clocks globally distributed in national metrological laboratories or astronomical institutes \cite{Panfilo2012,Panfilo2014}. The advantage of this virtual mean time scale (echelle atomique libre, EAL) is its superb stability and robustness. However, as EAL has been continuously free-running since 1977, its scale interval has an offset from the SI second of several parts in $10^{13}$. BIPM estimates the frequency of EAL with respect to all evaluations of PFSs and secondary frequency standards (SFSs) that are available at the time of TAI computation. The algorithm for deriving this estimation and its uncertainty are described in \cite{bipm1977}, and the result is published every month in Circular T as the ``duration of the TAI scale interval'' over the one-month interval of TAI computation.
%The prediction error was further evaluated by a postprocessing that predominantly refers the calibrations in the respective evaluation interval\cite{bipm1977}. This residual difference is commonly published in Circular T with an estimation interval of one month.

In contrast to the robustness of TAI, the operation of optical clocks seldom continues for one month, requiring TAI with a short averaging interval as the reference of frequency measurements. The shortest interval of the TAI scale is five days as TAI is computed by the BIPM only for 0:00 UTC every fifth day. Thus, optical frequency measurements normally utilize the TAI scale of five days as a reference. The calibration of the TAI relative to the SI second, on the other hand, is normally available for only one-month mean in Circular T. Therefore, we need to estimate possible error in assimilating the TAI frequency over five days to the one-month mean published in Circular T. This effect was called ``dead time (TAI)'' in the error budget, and was the most dominant source of uncertainty in the previous work \cite{jjap}.

In this report, we describe how to reduce this dominant uncertainty by employing the calibration of the TAI scale on each five-day interval of operation of our optical clock using the same algorithm as that described in \cite{bipm1977}.% By reevaluating the link part of our previous work \cite{jjap}, the frequency shifted for $3\times 10^{-16}$
With respect to our previous work \cite{jjap}, the re-evaluation of the link part has led to a $3\times 10^{-16} $correction of the $^{87}{\rm Sr}$ clock frequency
and the total uncertainty of the $^{87}{\rm Sr}$ clock frequency was reduced to less than $1\times 10^{-15}$ for the first time without local PFSs. The revised frequency is consistent with other  measurements employing local Cs fountains.

\section{Calibration of the TAI scale with estimation interval of five days}

\begin{table*}
\caption{Bias $\Delta$ and uncertainty $\delta$ of the TAI scale for a previous work \cite{jjap} and for this work. Instead of the estimation interval of 30 days (top row) which is routinely published by BIPM in Circular T, we calculated $\Delta$ and $\delta$ with an estimation interval of five days only for this work (second row) so as to match the estimation interval to the campaign length. The evaluation in the second row
incorporates all calibrations provided by all PFS within one year of our campaign. On the other hand, the evaluation in the third row considers only those PFSs which were active in our campaign duration. The difference between those indicates a rather minor impact on the results. }
\label{tab:1}       % Give a unique label
\begin{center}
\begin{tabular}{l|rr|rr|rr}
\hline\noalign{\smallskip}
campaign\# &\multicolumn{2}{|c|}{1} & \multicolumn{2}{|c|}{2} & \multicolumn{2}{|c}{3} \\
\noalign{\smallskip}\hline\noalign{\smallskip}
MJD & \multicolumn{2}{|c|}{57059-57064} & \multicolumn{2}{|c|}{57079-57084} & \multicolumn{2}{|c}{57124-57129} \\
& $\Delta$ & $\delta$ & $\Delta$& $\delta$ & $\Delta$ & $\delta$\\
\hline
TAI-SI second &&&&&& \\
(30 days, $\times 10^{-16}$)$^1$ & -4.4 & 2.6 & -2.7 & 2.5 & -2.3 & 2.6 \\
TAI-SI second &&&&&& \\
(5 days, $\times 10^{-16}$) & -2.5 & 8.1 & -6.8 & 8.2 & -4.5 & 7.4 \\
TAI-SI second only by PFS  &&&&&& \\
active over the interval (5 days, $\times 10^{-16}$) & -4.8 & 8.2 & -7.8 & 8.3 & -4.8 & 7.4 \\
\noalign{\smallskip}\hline
\multicolumn{7}{l}{$^1$ From Circular T No. 326, 327, 328}
\end{tabular}
\end{center}
\end{table*}

The least-squares method of deriving the consistent frequency error and the uncertainty of the TAI scale is described in \cite{bipm1977}. Here, the estimation interval is a free parameter, although BIPM regularly calculates only for an estimation interval of one month. %By estimating the TAI frequency over exact five-day intervals of operation of our optical clock, we optimize the evaluation of the uncertainty in the link to the SI second.
By setting the estimation interval to be 5 days, we no longer suffer from the discrepancy of the TAI interval between the interval of our optical frequency measurement and the interval for the calibration of the TAI with reference to the SI second. Thus, it is not required to introduce the uncertainty in regarding the two mean frequencies as being identical.

The new calibrations of TAI obtained with the estimation interval of five days are summarized in Table 1. There were three campaigns (\#1--3) for five (or four) days each. The maximum difference in the correction from the one-month average is $4\times 10^{-16}$ at maximum, which is consistent with the instability of TAI \cite{Petit}. When we use the five-day average, the link from TAI to a flywheel oscillator and that to the SI second have the same averaging interval. Thus, there is no additional dead-time error here. However, the uncertainty of the TAI scale calibration increases from about $2.6 \times 10^{-16}$ over one month to about $8\times  10^{-16}$ over five days, mostly as a result of the increased dead-time between the PFS evaluation intervals and the five-day estimation interval. Nevertheless, this dead-time uncertainty is more correctly estimated using the algorithm in \cite{bipm1977}, and thus the final uncertainty is better when using five-day intervals.

 In order to evaluate the weighted average of the three campaigns, we need to investigate the correlation among these three TAI-SI second calibrations. For instance, the systematic error of Cs fountains is totally correlated when identical fountains are operated in three campaigns. Table 2 shows the PFSs that were operated in the three campaigns. The uncertainties in the calibration include the type-A and type-B uncertainties of the PFS, $u_A$ and $u_B$, respectively, as well as the link uncertainty $u_l$. The type-A uncertainty is a statistical uncertainty determined by the measurement duration and the magnitude of the white frequency noise in the PFS system. The type-B uncertainty is the uncertainty of the systematic bias. In addition, the calibration by a SFS has an uncertainty of the secondary representation of the second $(u_{Srep})$. These uncertainties are shown in Circular T together with the total uncertainty $u$. It is reasonable to regard $u^{-2}$ as proportional to the weight that determines the TAI-SI second calibration \cite{bipm1977}. The weight also allows us to estimate how much of $u$ is attributed to the systematic part, namely $u_B$. Given the systematic uncertainty of the PFS $i$ to be $u_{B,i}$, the weighted mean of the systematic part $\overline{u_B}$ in the campaign $\# j$ is written as
 \begin{equation}
 \overline{u_{B,j}} = \left( \sum_i \left( w_{i,j} u_{B,i}\right) ^2 \right)^{1/2},
 \end{equation}
 where $w_{i,j}$ is the weight of the calibration by PFS $i$ in the campaign $\# j$.
 $\overline{u_{B,j}}$ for each campaign resulted in $(1.6 - 1.7) \times 10^{-16}$, indicating rather minor contribution comparing with typical $u$ of several parts per $10^{16}$.

 The weights derived from $u^{-2}$ suggest that the calibrations in the three campaigns rely on SYRTE-FO2 with the largest weight of 44\%. While systematic uncertainties among PFSs should have no correlations, this rather high occupancy of a single fountain prevents us from employing the reduced systematic uncertainty by the statistical mean, therefore we consider a systematic part of $1.6\times 10^{-16}$ also for the mean of the three determinations. The rest of uncertainties of the campaign $\#j$, $u_{\rm{random}, j}$ is calculated to be
\begin{equation}
u_{{\rm random}, j} = \left( \sum_{i=1}\left( u_{A, i, j}^2+u_{l, i, j}^2 \right) \right)^{1/2},
\end{equation}
where $u_{A,i,j}$ and $u_{l,i,j}$ are the type-A and link uncertainties of the PFS $i$ in campaign $\#j$, respectively. %$u_{Srep}$ is mostly zero because this is applied only for SFS.
These random parts are averaged to obtain the total random error of the three campaigns, which resulted in a value of $5.6\times 10^{-16}$. Considering the systematic part of $1.6\times 10^{-16}$, we concluded that the total uncertainty of the averaged TAI scale against the SI second is $5.8\times 10^{-16}$.

\begin{table*}
\caption{List of primary frequency standards (PFSs) and a secondary frequency standard (SFS) that contributed to the calibration of the TAI scale of the three campaigns. The effective weight of the PFS (or SFS) is proportional to the $u^{-2}$, where $u$ is the total uncertainty in the calibration of the TAI scale by the respective PFS or SFS. Note that the $u$ of the SFS here includes $u_{Srep}$.  The weights of the campaigns (\#1, \#2, and \#3) were 16.4, 65.8, and 17.8\%, respectively, according to the statistical uncertainties of the previous measurement. \cite{jjap} }
\begin{center}
\begin{tabular}{l|rr|rr|rr|r}
\hline\noalign{\smallskip}
campaign\# &\multicolumn{2}{|c|}{1} &\multicolumn{2}{|c|}{2} & \multicolumn{2}{|c|}{3} &  Total \\
& weight & operation & weight & operation & weight & operation & weight \\
& (\%)       & (days) & (\%)& (days)& (\%)& (days) & (\%)\\
\hline
PTB-CS1 & $<1$ & 30 & $<1$ & 30 & $<1$ & 30 & $<1$ \\
PTB-CS2 & $<1$ & 30 & $<1$ & 30 & $<1$ & 30 & $<1$ \\
IT-CsF2 & 16& 20 & && 10 & 15&  4\\
NIM5 & & & 3 & 25 & 3 & 20 & 3 \\
NIST-F2 & & & 26 & 20 &  &  & 17 \\
PTB-CSF1 & & & & & 14 & 30& 2 \\
PTB-CSF2 & & & & & 45 & 15& 7 \\
SU-CsF02 & 17& 30 & 14& 25 & 16 & 35 & 15 \\
SYRTE-FO1& 27& 30 & &&  && 5\\
SYRTE-FO2& 36& 25 & 52& 30& 24 & 25 & 44 \\
SYRTE-FORb& 4& 30 & 5& 30 & 3 & 30& 4 \\
\noalign{\smallskip}\hline\noalign{\smallskip}
weighted mean of $u_B$ &&&&&&& \\
 ($\overline{u_B}$, $\times 10^{-16}$) & & 1.6 & & 1.6 & & 1.7 & 1.6 \\
weighted mean of other &&&&&&& \\
uncertainties $((\delta^2-\overline{u_B}^2)^{1/2}, \times 10^{-16})$ & & 7.9 & & 8.0 & & 7.2 & 5.6 \\
\noalign{\smallskip}\hline\noalign{\smallskip}
Total uncertainty $\delta$ &&&&&&& \\
(From second row of Table 1, $\times 10^{-16}$) & & 8.1 & & 8.2& & 7.4 & 5.8
%\multicolumn{8}{l}{$^1$ Operation of 10 days in the previous month.}
\end{tabular}
\end{center}
\end{table*}

In Circular T, the calibration of the TAI scale is determined not only by the PFSs that were active on the specific month but also by other PFSs that were operated in the past. Prior to the invention of the atomic fountain, the fractional inaccuracies of the PFSs were at the same level or larger than the long-term instability of the TAI scale. Thus, the evaluations using all contributing PFSs definitely helped to estimate the TAI scale. After the Cs fountains reached an accuracy better than the long-term instability of the TAI scale, however, the contribution of the PFSs that were operated outside the estimation interval has decreased. Therefore, we also determined the five-day calibration using only the PFSs that were active on the five days of each campaign. The result is shown in the bottom row of Table 1, where the maximum difference from the calibration by all PFSs is $2\times 10^{-16}$ , and the uncertainty does not change at all. Since we cannot find a significant difference between the two calibrations, the normal calibration including all PFSs is employed in the following discussion.

\begin{table*}
\caption{Uncertainty budget of the previous evaluation \cite{jjap} and this revised evaluation. The dead-time uncertainty of TAI becomes zero. On the other hand, the uncertainty of the TAI calibration increased owing to the reduction in the estimation interval relative to the estimation intervals reported from PFSs.  }
\begin{center}
\begin{tabular}{l|rr|rr|}
& \multicolumn{2}{|c|}{campaign \#3 $(\times 10^{-17})$} & \multicolumn{2}{|c|}{total $(\times 10^{-17})$} \\
contributor & \cite{jjap}& this work  & \cite{jjap} & this work\\
\noalign{\smallskip}\hline\noalign{\smallskip}
%statistical& 9 & $\leftarrow$ & 19 & 11 \\
statistical& 9 & $\leftarrow$ & 19 & $\leftarrow$ \\
strontium & 9 & $\leftarrow$ & 10 & $\leftarrow$ \\
gravity & 11 & $\leftarrow$ & 8 & $\leftarrow$ \\
dead time (HM4) & 27 & $\leftarrow$ & 19 & $\leftarrow$ \\
HM4-UTC(NICT) & 5 & $\leftarrow$ & 4 & $\leftarrow$ \\
UTC(NICT)-TAI & 98 & $\leftarrow$ & 69 & $\leftarrow$ \\
dead time (TAI) & 110 & 0 & 76 & 0 \\
TAI-SI second & 26 & 74 & 25 & 58 \\
\noalign{\smallskip}\hline\noalign{\smallskip}
Total & 155 & 127 & 110 & 95
\end{tabular}
\end{center}
\end{table*}

The uncertainty budget with the revised link uncertainty is shown in Table 3 along with the previous one \cite{jjap}. Changes can be seen in the bottom two rows, namely, dead time (TAI) and TAI-SI second. The dead-time uncertainty in assimilating the TAI frequency over five days to that computed by the BIPM over 30 days was removed with a penalty of a larger TAI-SI second uncertainty for each five-day evaluation. Nevertheless, the trade-off is beneficial, and the overall uncertainty was $9.5\times 10^{-16}$, which is below $1\times 10^{-15}$ for the first time  for a measurement using the TAI.
%The difference between the two intervals boosts the residual uncertainty of the least square method \cite{bipm1977}. In addition, the statistical uncertainty was reduced because we found error in the previous work, where we evaluated the statistical error from the scattering of three mean frequencies of the campaigns. However, part of the reason of this scattering is already listed up in other row such as dead time (HM4) and UTC(NICT)-TAI link.
%The 5-day average of a hydrogen maser frequency is derived from the linear fitting to 5 frequencies. The weighted average of this fitting error was $1.0\times 10^{-16}$, which is listed as the statistical uncertainty.

\begin{figure}
\centerline{\includegraphics[width=\columnwidth]{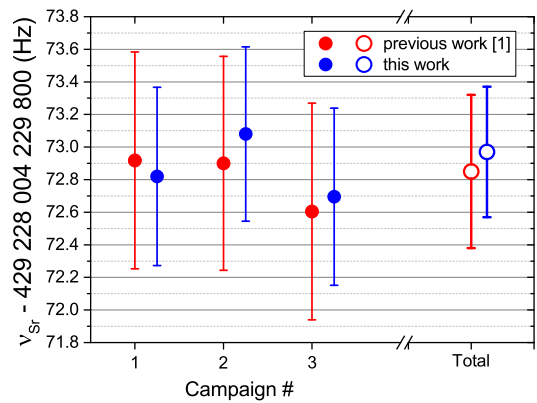}}
\caption{Absolute frequencies of the campaigns obtained using the improved frequency link (blue). Red points indicate the results of the previous work \cite{jjap}. The total weighted means based on the statistical uncertainty are also shown, where the value shifted by up to three parts per $10^{16}$ and the uncertainty was reduced to less than $1\times 10^{-15}$. }
\end{figure}

Finally, the revised absolute frequencies measured in the three campaigns are summarized in Fig. 1 together with the previous ones. For a given 5-days interval, the maximum difference between the two frequency links is 0.18 Hz $(4\times 10^{-16}$ (see Table 1). The calculation of the weighted average of the three campaigns resulted in a final value of 429 228 004 229 872.97 (40) Hz using the novel frequency link. We consider that the frequency of the measurement in 2015 \cite{jjap} should be modified to this new value because the frequency reevaluated here has a lower uncertainty. This revised frequency agrees with other measurements recently performed in various laboratories. The frequencies are compared in Fig. 2, where the first six points were additionally taken into consideration for the determination of the recommended standard frequency of the CIPM 2015 ( Comit\'{e} international des poids et mesures 2015). The red circle \cite{jjap} was reevaluated in this work as the blue circle. The last four points were reported after the CCTF2015 (Consultative Committee for Time and Frequency 2015). In particular, the degree of agreement is excellent with the results recently reported from SYRTE \cite{SYRTENew} and PTB \cite{PTBNew} using locally available state-of-the-art fountains.

\begin{figure}
\centerline{\includegraphics[width=\columnwidth]{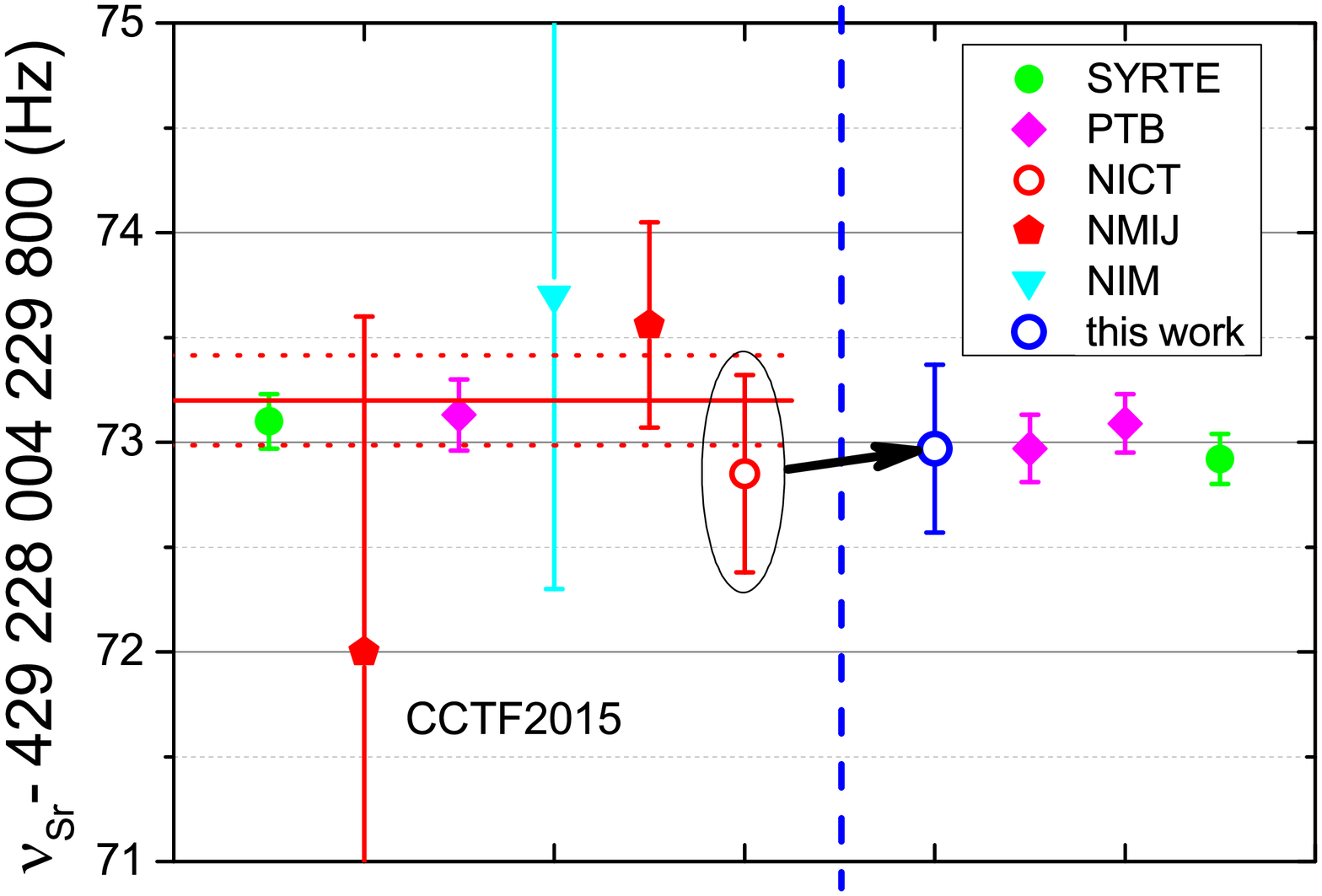}}
%\resizebox{0.3\hsize}{!}{
%\includegraphics*{fig1.eps}
%\vspace{5cm}       % Give the correct figure height in cm
\caption{$^{87}{\rm Sr}$ clock transition frequencies recently reported from various institutes including SYRTE \cite{SYRTENew,SYRTE2013}, PTB \cite{PTBNew,PTB2014}, NICT \cite{jjap}, NMIJ \cite{NMIJ1,NMIJ2}, and NIM \cite{NIM}. The previous result (red empty circle) was reevaluated to the blue empty circle in this work.}
\end{figure}

\section{Summary}
In summary, we reduced the uncertainty of the previous absolute frequency measurement \cite{jjap} by employing a TAI estimation interval of five days, which matched the length of the measurement campaign. Temporally distributed operation of optical clocks over the five-day TAI grid led to the suppression of the link uncertainty between local flywheel oscillator and TAI in previous work. The uncertainty of the frequency  link from TAI to the SI second over the 5-day measurement intervals was also reduced in this work, resulting in a total uncertainty below $1\times 10^{-15}$ for the first time for a TAI-based measurement.

Note that further improvement would result from a longer operation interval as this would improve the uncertainty in both the frequency of TAI and the frequency transfer from UTC(NICT) to TAI. Another potential advantage of the TAI-based measurement is that the limitation due to the possible systematic bias of a specific PFS is mitigated by the contribution of other PFSs. For instance, estimating the frequency simultaneously with reference to four similarly accurate PFSs via TAI statistically reduces the PFS-originated systematic uncertainty by a factor of two. Regular calibration of the TAI scale using optical clocks will not only contribute to the maintenance of the time scale, but also contribute to the determination of accurate standard frequencies.

\begin{acknowledgements}
  The authors are grateful for the discussion with Y. Hanado, F. Nakagawa, H. Ito, T. Gotoh, M. Kumagai, and M. Hosokawa. It is also worth noting that the measurement depends on Cs fountains at National Metrological Institutes provided for the calibration of the TAI scale.
\end{acknowledgements}

\end{document}